\documentclass[twocolumn,showpacs,preprintnumbers,amsmath,amssymb]{revtex4}

\usepackage{graphicx}
\usepackage{dcolumn}
\usepackage{bm}

\begin{document}


\title{Robustness of non-abelian holonomic quantum gates against parametric noise}
\author{Paolo Solinas,$^{1}$ Paolo Zanardi,$^{2, 3}$ Nino Zangh\`{\i}$^{1}$}

\affiliation{
$^1$ Istituto Nazionale di Fisica Nucleare (INFN) and
Dipartimento di Fisica, Universit\`a di Genova,
Via Dodecaneso 33, 16146 Genova, Italy \\
$^2$ Department of Mechanical Engineering,
Massachusetts Institute of Technology, Cambridge Massachusetts 02139\\
$3$ Institute for Scientific Interchange (ISI),
Viale Settimio Severo 65, 10133 Torino, Italy}  

\date{\today}

\begin{abstract}
We present a numerical study of the robustness of 
a specific class of non-abelian holonomic quantum gates . 
We take into account the parametric noise due
to stochastic fluctuations of the control fields which drive the
time-dependent Hamiltonian along an adiabatic loop.  The performance
estimator used is the state {\it fidelity} between noiseless and noisy holonomic gates.
  We carry over our analysis with different correlation times 
and  we find out that noisy holonomic gates seem to be close to the noiseless
ones for 'short' and 'long' noise correlation times.  This result can
be interpreted as a consequence of to the geometric nature of the
holonomic operator. 
Our simulation have been  performed by using  parameters relevant
to the  excitonic proposal for implementation of holonomic quantum computation
 [P. Solinas {\it et al.} Phys. Rev. B {\bf 67}, 121307 (2003)]
\end{abstract}

\maketitle
\section{Introduction}

The use of uniquely quantum  phenomena to process information has led to surprising results
in quantum key distributions \cite{crypto},  information transfer protocols 
\cite{teleportation} and computation \cite{algos}.
From the point of view of the actual
implementation of these theoretical protocols 
a main challenge is posed by the fact that generically quantum states are very delicate objects
quite  difficult to control with the required accuracy. 
The interaction with  the many degree of freedom of the environment causes  
the loss of information (decoherence) and  moreover errors in processing the 
information lead to a wrong output state (control errors). 
The first problem has been extensively studied over the past few years 
and a few ways  to overcome it have been proposed and experimentally realized.
These strategies include error avoiding \cite{error_avoiding}, 
error correcting strategies \cite{error_correcting} and  
decoupling techniques \cite{bang-bang}.

A new approach called {\it topological quantum computation} has been argued
to be able to  effectively solve both of them and open new ways to inherently
robust quantum computation \cite{topological}.
Information is encoded in topological degrees of freedom of a system 
which are not sensitive to the local environment-noise effects and then are 
robust against decoherence \cite{tns}.
Unfortunately to the date, no simple feasible physical system has been 
identified to this aim; in fact, the systems proposed are usually complicated 
many-particle ones living over 
a macroscopic non-trivial structure (e.g. torus or cylinder topology).
On the other hand, we can develop {\it topological information processing}, 
where the operator used depends on topological controls and then are 
robust against a the unwanted fluctuations of the  driving fields.
In this case, a important intermediate step is the {\it geometrical 
quantum computation} and particularly promising is the fully-geometrical 
approach called Holonomic Quantum Computation (HQC) \cite{HQC}.

At variance with  topological information processing, for geometrical QC 
several implementation proposals have been made;  indeed the holonomic 
structure shows up  in a variety of quantum systems, both in its  Abelian 
(Berry) \cite{abelian} and non-Abelian version 
\cite{HQC_proposal,duan,paper1,long_pap}.
For this reason interest around HQC has recently 
growth leading also to proposals in which  the adiabatic request can be 
relaxed and  for implementing geometrical non-adiabatic quantum computation
\cite{non_adiabatic}.
One of the apparent advantages of HQC is that the 
gating time for holonomic gates does not depends on the logical 
operator applied but only on the adiabatic request; then
HQC may  lead to a new approach to  implement  
complex operators difficult to construct with the standard dynamical gates
(as discussed in \cite{long_pap}).
In view of its geometric nature i.e.,
dependence on {\em areas} spanned by loops in the control parameter manifold, 
HQC  has been suggested to be robust against  some class of errors 
\cite{preskill,ellinas}.
Nevertheless thorough studies aimed to address this important issue
are still relatively few and certainly not exhaustive \cite{HQC_noise}.

In this paper we will deal with the noise due to imprecise  
control of the system parameters during the evolution.
This error source will be referred to as   {\it parametric noise}.
We will show that the {\it fidelity} of the holonomic gates 
displays  three regimes, depending on the {\it ratio} between the adiabatic 
time and the noise correlation time.
This results can be understood in view of the geometrical dependence 
of the holonomic operator.
We will study in detail the class of holonomic gate proposed in \cite{paper1}
 in which the physical system are semiconductor quantum dots, the logical 
qubits are excitonic quantum state controlled by {\it ultrafast} lasers.

In section \ref{sec:model}, after a brief review of the holonomic approach, we describe the system used and the 
logical gate studied; moreover, it is discussed how we model the noise in the control 
parameters. In Sec. \ref{sec:simulation} we give a description of our 
simulations and show the results with different kind of noise processes
for two single qubit gates and for a two qubit gate.
A comparison with dynamical gates subject to the same 
noise is given too. Section \ref{sec:conclusions} contains the conclusions.

\section{Holonomic quantum gates with parametric noise}
\label{sec:model}

Let us consider  a family $\mathcal{F}$  of isodegenerate
Hamiltonians $H(\lambda)$ depending
on $m$ dynamically controllable parameters $\lambda$, in the HQC approach \cite{HQC}
one  encodes the information in a $n-$fold degenerate eigenspace $\mathcal{E}$ of an Hamiltonian
$H(\lambda_0)$.
Varying the $\lambda$'s and driving {\em adiabatically} $H(\lambda)$ along a 
loop in the  $\lambda$ manifold we produce a non-trivial transformation of the 
initial state $|\psi_0 \rangle \rightarrow U |\psi_0 \rangle$.
These transformations,
known as  {\it holonomies}, are the generalization of Berry's phase and
can be computed in terms of the Wilczek-Zee gauge connection
\cite{W-Z}:  $U(C)={\bf P} exp(\oint_C A)$ where $C$ is the loop
in the parameter space and $A=\sum_{\mu=1}^m A_{\mu} d\lambda_{\mu}$
is the $u(n)-$valued connection.
If $|D_i(\lambda)\rangle $ ($i=1,...,n$) are the instantaneous
eigenstates of $H(\lambda)$, the connection is
$(A_\mu)_{\alpha\beta}=\langle D_\alpha|{\partial}/{\partial\Omega^\mu}|
D_\beta\rangle$ ($\alpha$, $\beta=1,...,n$).
The set of holonomies associate with a given connection is known to be a subgroup of the group
of all possible $n$-dimensional unitary transformations;
when the dimension of this holonomy group and 
coincides with the dimension of $U(n)$ one is  able to perform universal 
quantum computation with holonomies \cite{HQC}.

For concreteness in this paper we will focus on  the class of  
holonomic quantum gates analyzed  in \cite{paper1,long_pap}. 
Logical qubits are given by polarized  excitonic states controlled by 
femtosecond laser pulses.
The parameters we have used in performing our simulations are those relevant
to this specific kind of physical systems.

First we concentrate  to  {\em one-qubit} gates.
Despite this might look at first as major limitation, we observe that, as far as  the holonomic structure
is concerned, the two-qubit gates are very similar. So we expect that most of the results
we are going to present here e.g., the existence of separate regimes,
should, to a large extent, hold true for two-qubit gates too.

The time-dependent interaction Hamiltonian in the interaction picture is 
\begin{eqnarray}
  H_{int} & = &
 - \hbar (\Omega_+|E^{+}_L\rangle +
   \Omega_-|E^{-}_L\rangle +
   \Omega_0|E^{0}_L\rangle )\langle G| \nonumber \\
   & + & h.c.
   \label{eq:ham_ec}
\end{eqnarray}
where $|E^i_L\rangle$ ($i=+,-,0$) are the polarized excitonic states 
(two logical and one {\it ancilla}) and $|G\rangle$ is the ground 
state (absence of exciton). 
This Hamiltonian family admits two {\em dark states} i.e.,
$H_{int}(\Omega)  |D_i(\Omega)\rangle=0,\,(i=1,2).$
This two-fold degenerate manifold represent contains our encoded logical qubit: 
$|0\rangle_L:=|E^{+}_L\rangle,\;|1\rangle_L:=|E^{-}_L\rangle.$
In Refs.  \cite{duan,paper1} it has been shown that the Wilczek-Zee connection associated
to the Hamiltonian family (\ref{eq:ham_ec}) allows to construct  universal one-qubit
gates. These are realized by giving an explicit prescription for driving
the control parameter  $\Omega$'s along suitable adiabatic loops.

We suppose now to  add to the control $\Omega$-field
a 'small' noise which perturbs our trajectory on the control manifold.
We test the robustness of the geometrical {\it mixing } single-qubit gate proposed
in Ref. \cite{paper1}. To obtain this gate, we made the following loop
in the parameter space :
$\Omega_-(t) = \Omega~ \sin\theta \cos\varphi $,
$\Omega_+(t) = \Omega~ \sin\theta \sin\varphi $ and 
$\Omega_0(t) = \Omega~ \cos\theta $ 
(where the $\Omega$ parameters is fixed and the $\theta (t)$ and 
$\varphi (t)$ are time dependent)
and the holonomic operator obtained 
at the end of the loop is $U = e^{i \phi \sigma_y}$, 
(where $i \sigma_y = |E^+_L \rangle \langle E^-_L| - |E^-_L \rangle 
\langle E^+_L|$).
The geometrical parameter $\phi = \oint \sin\theta d\theta d\varphi$  
is the solid angle spanned by the parameter vector 
$\vec{\Omega} = (\Omega_+ , \Omega_- , \Omega_0)$
on the parameter manifold (sphere).
Changing the relation between $\theta$ and $\varphi$ we change the loop 
and then the value of $\phi$. 
We choose a loop in order to obtain $\phi = \pi/2$ and 
$U = exp(i \pi \sigma_y / 2) = 
|E^+_L \rangle \langle E^-_L| - |E^-_L \rangle \langle E^+_L| $.

The logical operator $U$ depends only on geometrical parameters 
(i.e. solid angle swept on the parameters manifold), every perturbation
that changes the trajectory in the control manifold changes the operator
leading to computational errors. This notwithstanding,
the perturbations leaving (almost) unchanged this angle will not affect the 
holonomic operator.
Then we expect that even strong fluctuations - 
provided their time scale is sufficiently fast - average out,
leaving in this way the angle unchanged and thus 
not affecting the computation.

The control parameters are the intensities and the phases of the lasers 
(since we suppose to be in resonant condition, we have fixed frequencies)
perturbed by external noise (one for the phases and one for the 
intensities).
To clarify which of these mainly affects the gate operation,
first we separate the two types of errors (intensity and 
phase fluctuations)  and then we apply both of them.

We use a straightforward model for the noise :
we extract a random number from a probability distribution, we add 
a constant noisy field for the time $T_n$
to the evolving control field, then we extract another random number and so on.
To simplify the simulations we choose $T_n$ in such a way that $T_{ad}$ 
is a multiple of $T_n$ ($T_{ad} = n_r T_n$).
The fundamental parameter is the {\it noise time} $T_n$~ that is the lapse of 
time of each random extraction; i.e. it represents the time scale of each 
random fluctuation.

\begin{figure}[t]
  \begin{center}
    \includegraphics[height=5cm]{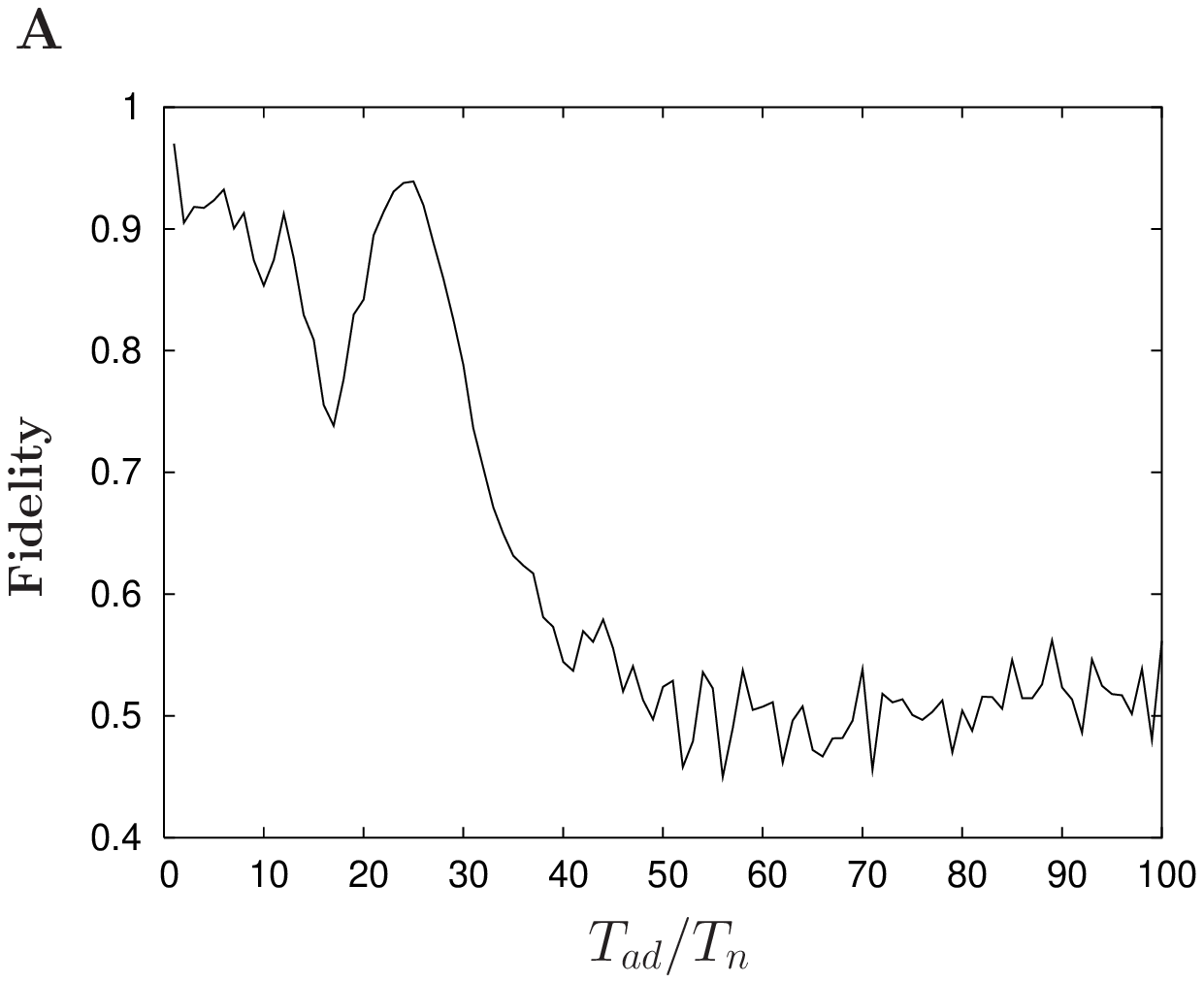}
    \includegraphics[height=5cm]{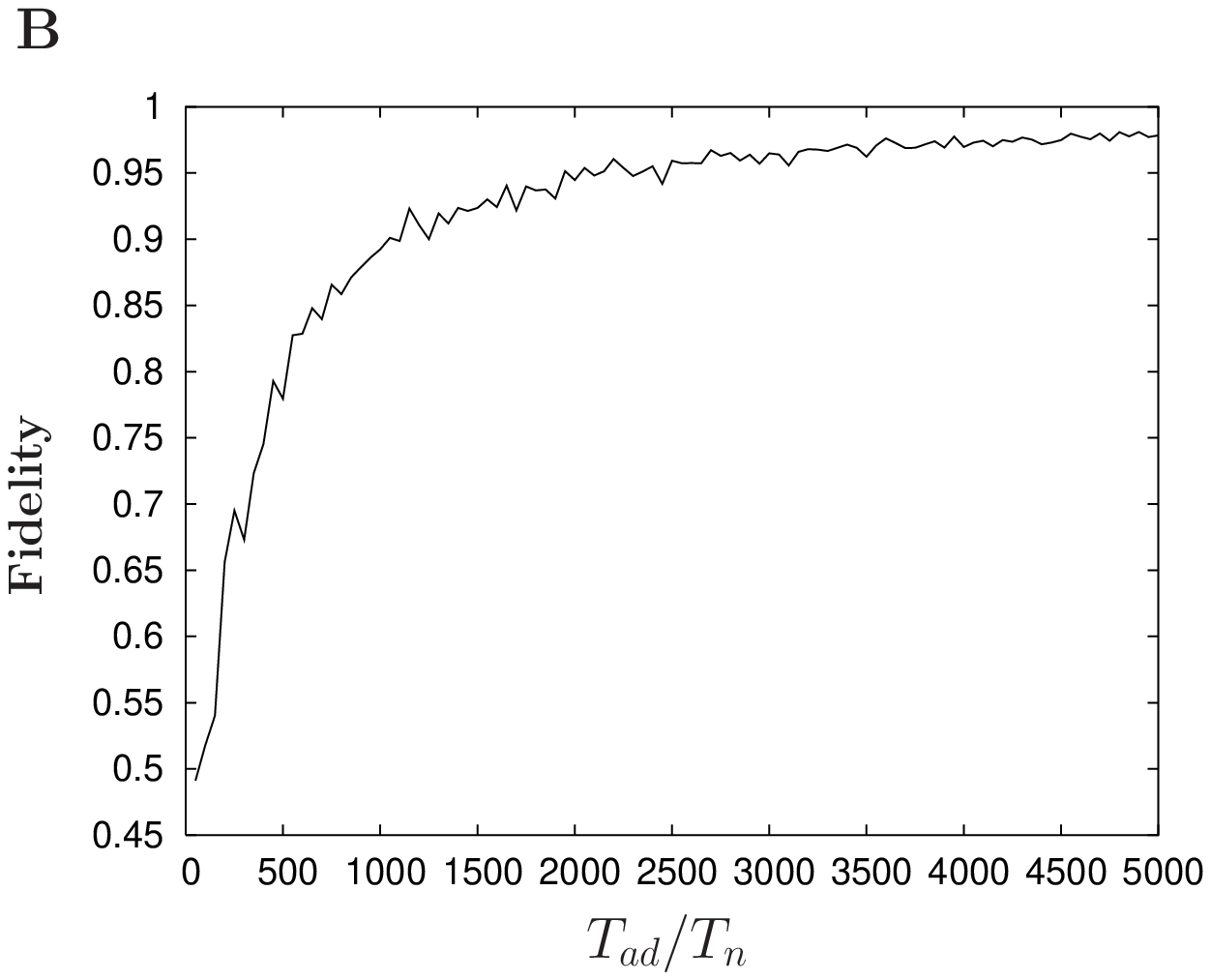}
    \caption{\label{fig:ad_int02_150} {\it Fidelity} for holonomic gate with 
      $\Omega^{-1} = 50~fs$, $T_{ad}=7.5~ps$ ($\Omega~T_{ad}=150$)
      and $\delta \Omega = 0.1 \Omega$. (A) Slowly varying fluctuations
      and (B) fast varying fluctuations.}
  \end{center}
\end{figure}

For the 'intensity' noise we modify only the value of the Rabi frequencies
$\Omega$. We have three lasers turned on and we suppose they 
have independent fluctuations $\delta\Omega_i$ 
(i.e. every $T_n$~ we extract three random numbers).
The evolution on the control manifold is described by:  

\begin{equation}
  \left \{
  \begin{array}{lcl}
    \Omega_-(t) &=& \Omega~ \sin\theta \cos\varphi + \delta \Omega_-(t) \nonumber \\
    \Omega_+(t) &=& \Omega~ \sin\theta \sin\varphi + \delta \Omega_+(t) \nonumber \\
    \Omega_0(t) &=& \Omega~ \cos\theta + \delta \Omega_0(t)
  \end{array}
  \right.
  \label{eq:int_noise}
\end{equation}

In this case, the Rabi frequencies remain real parameters, but, if we 
introduce a 'phase' noise ($\Omega_j \rightarrow e^{i \xi_j} \Omega_j$),
they acquire an imaginary part.
The random numbers are taken as before and the evolution  in the control 
manifold is: 

\begin{figure}[t]
  \begin{center}
    \includegraphics[height=4.7cm]{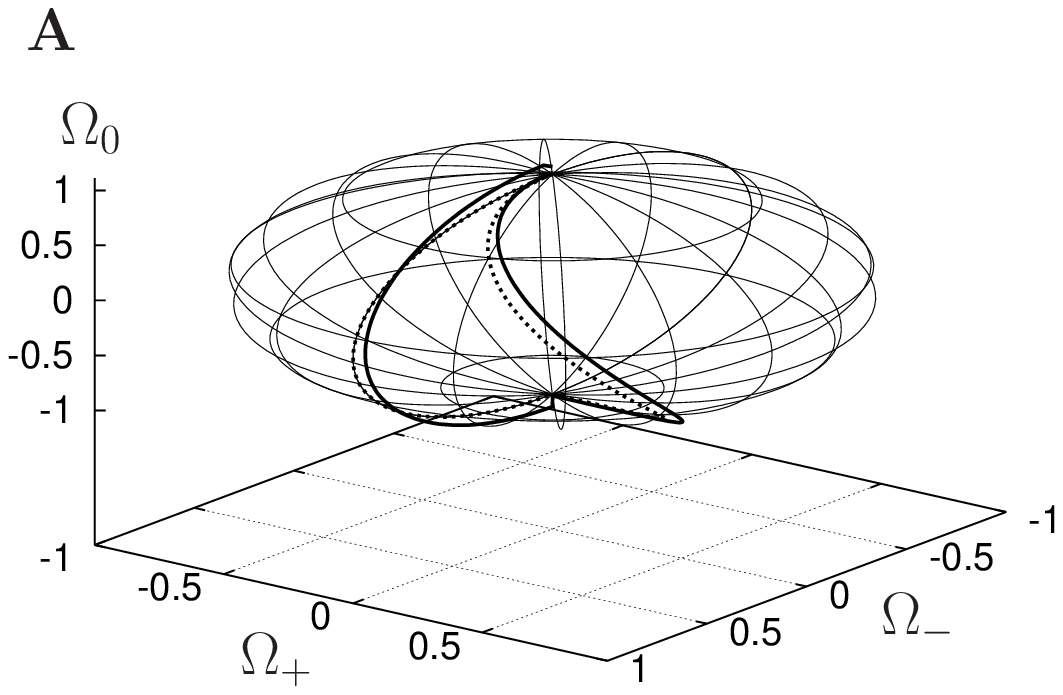}
    \includegraphics[height=4.7cm]{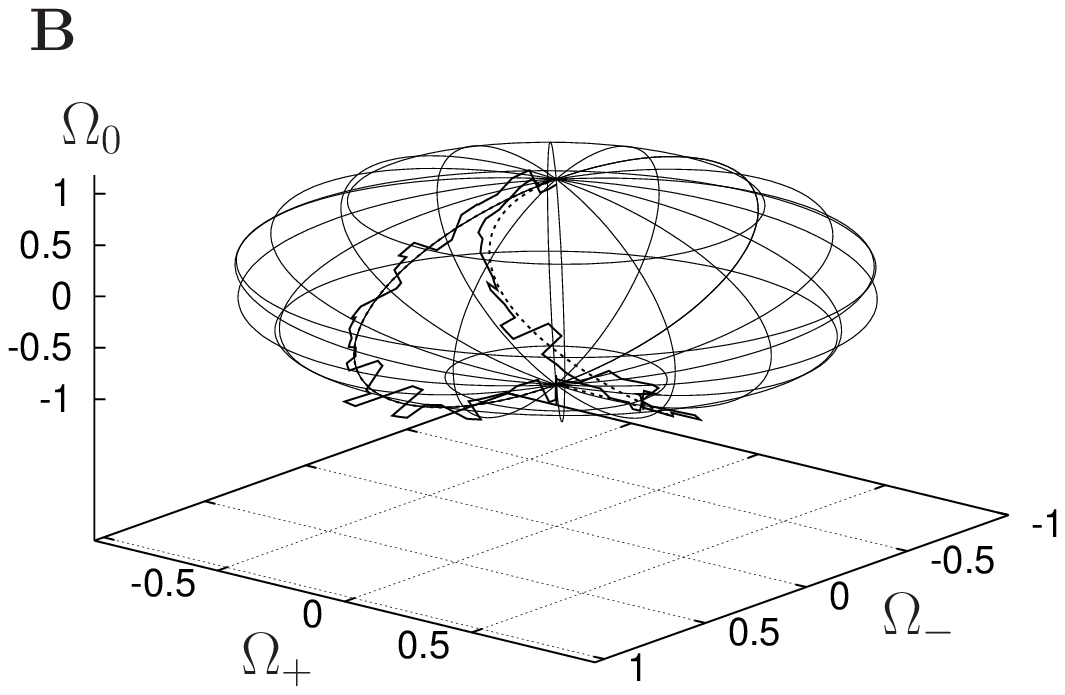}
    \caption{\label{fig:par_loop} Loop in the parameter space for the 
      holonomic gate with (solid line) and without noise (dashed line).
      With extraction of 2 ($T_n = T_{ad}/2$) (A) and 
      70 random numbers ($T_n = T_{ad}/70$) (B).
      The sphere radius is normalized to $\Omega=1$.}
  \end{center}
\end{figure}

\begin{equation}
  \left \{
  \begin{array}{lcl}
    \Omega_-(t) &=& e^{i\xi_-}~\Omega~ \sin\theta \cos\varphi \nonumber \\
    \Omega_+(t) &=& e^{i\xi_+}~\Omega~ \sin\theta \sin\varphi \nonumber \\
    \Omega_0(t) &=& e^{i\xi_0}~\Omega~ \cos\theta
  \end{array}
  \right.
  \label{eq:phase_noise}
\end{equation}

The most general and complicated situation is when  
both 'intensity' and 'phase' noise are present. 

\section{Simulations}
\label{sec:simulation}

For all the simulation we choose the parameters used in Refs. 
\cite{paper1,long_pap}
which satisfy the adiabatic condition :
$\Omega= 0.02$~fs$^{-1}$, $T_{ad}= 7.5$~ps and $\Omega~T_{ad} =150$. 
The probability distribution for the noise is 
a Gaussian with zero mean and $<\sigma>=< \frac{\delta \Omega}{\Omega} >=
0.1$ where $\delta \Omega$ is the fluctuation of the Rabi frequency.
Though this value is far below the experimental control, we use it to 
understand the robustness of holonomic gates against strong perturbations.

Once the evolution of the state with the noise  has been computed, we must compare it 
with the  ideal one in which  the noise is absent.
In order to make such a comparison quantitative we introduce the { \it fidelity}  
\begin{equation}
\mathcal{F} = \sqrt{ \langle \psi^{id}_{out}|
\rho_{out} |\psi^{id}_{out} \rangle}
\label{fid}
\end{equation}
 where $|\psi^{id}_{out} \rangle$ is the 
final state without noise and $\rho_{out}$ is the density matrix of 
associated to the noisy final state.
In our case the evolution is unitary, then 
$\rho_{out} = |\psi_{out}^{noise} \rangle \langle \psi_{out}^{noise}|$ 
and the {\it fidelity} reduce to be a scalar product between the noisy and 
the ideal state.
To eliminate the dependence of (\ref{fid}) on the initial states we make a 
sampling of the initial state space and then average the results.
Even if we have a four dimensional working space, 
the initial state space has dimension two; in fact, in the ideal gates 
(once satisfied the adiabatic condition) we always start and end in a 
superposition of the logical states $|E^+_L\rangle - |E^-_L\rangle$.
This simplifies the sampling procedure because we can take the Bloch sphere as
sampling space.
We sample the Bloch sphere with $18$ states \cite{sample} and for each of them 
calculate the {\it fidelity}.
Fixed the random numbers extracted during the evolution (i.e. fixed $T_n$),
we have also to take into account the dependence from the series of the 
random number extracted.
For every sampled state we make five different realization with different 
extracted random numbers and calculate 
the average {\it fidelity} for the particular initial state.
Finally, we make the average {\it fidelity} for all the sampled states to 
obtain the {\it fidelity} of our gate for a fixed $T_n$.

In the Figures presented here on the $y$~ axis is plotted the {\it fidelity}
and on the $x$~axis is plotted the number of extractions (that is the 
the {\it ratio} $T_{ad}/T_n$ between the adiabatic and {\it noise time}).

\begin{figure}[t]
  \begin{center}
    \includegraphics[height=5.3cm]{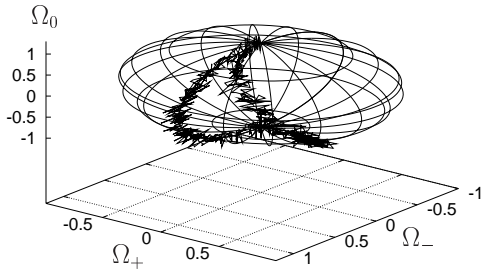}
    \caption{\label{fig:short_noise}  Loop in the parameter space for the 
      holonomic gate with noise 
      and the extraction of 5000 random numbers ($T_n = T_{ad}/5000$).
      The sphere radius is normalized to $\Omega=1$.}
  \end{center}
\end{figure}

In Figures \ref{fig:ad_int02_150} we report the simulated  evolution for two 
regimes of noise. In fig. \ref{fig:ad_int02_150} (A) we plot the 
{ \it fidelity} when we extract up to $100$ random numbers during the evolution 
($T_{ad}/100 \le T_n \le T_{ad}$).
Up to $30$ random numbers the average { \it fidelity} is $0.875$, 
while it decreases up to a minimum of about $0.5$ (with total average of 
$0.632$) if we extract more random numbers.

A possible interpretation of this effect can be given looking at
Figure \ref{fig:par_loop} where we show the evolution on the parameter sphere.
In figure \ref{fig:par_loop} (A) $n_r=2$~ ($T_n = T_{ad}/2$) 
and we change the noise field twice during the evolution.
Despite the intense noise, the shape of the loop is still clearly visible, 
it is simply shifted with respect to the ideal one. 
The value of the resulting solid angle swept is near to the ideal one.
In figure \ref{fig:par_loop} (B) we extract $n_r=70$~ random extractions 
during the adiabatic evolution ($T_n = T_{ad}/70$).
The fluctuations are too intense and too few to cancel out then, 
as we expected, the solid angles, swept, respectively during the ideal 
and the noisy loops, are different.
This can explain the result in figure \ref{fig:ad_int02_150} (A).

As stated before, even if we have strong fluctuations, we expect that if we 
extract  many random numbers the noise in average does not affect the solid 
angle and leave the holonomic operator unaffected.
This is what seems to be confirmed by the simulations illustrated in Figure 
\ref{fig:ad_int02_150} (B) where 
we extract from $50$ to $5000$ random numbers.
The {\it fidelity} increases and it is even better of those in 
figure \ref{fig:ad_int02_150} (A); the average {\it fidelity} is $0.918$ 
but it increases to $0.956$ if take into account the values from $1000$ to 
$5000$ random extractions.
The relative loop in the parameter space is shown in figure 
(\ref{fig:short_noise}) where we extract $5000$ random numbers 
during the evolution: the fluctuations have $\delta \Omega = 0.1~ \Omega$ 
but are so quick that cancel out.

From these simulations it is evident that we have three regimes in our 
model which can be explained as the sign of the geometrical
dependence of the holonomic operator:
\begin{figure}[t]
  \begin{center}
    \includegraphics[height=5cm]{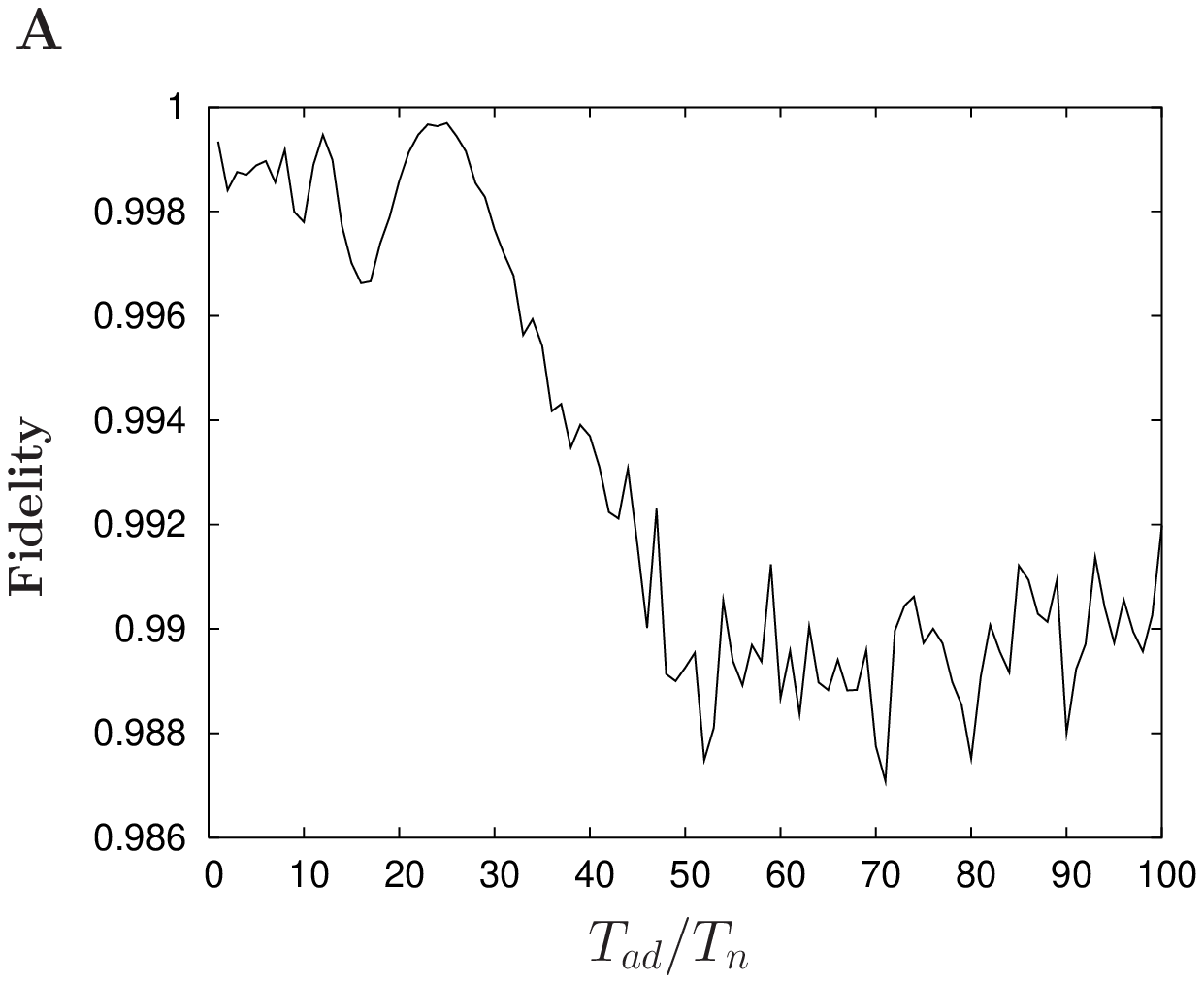}
    \includegraphics[height=5cm]{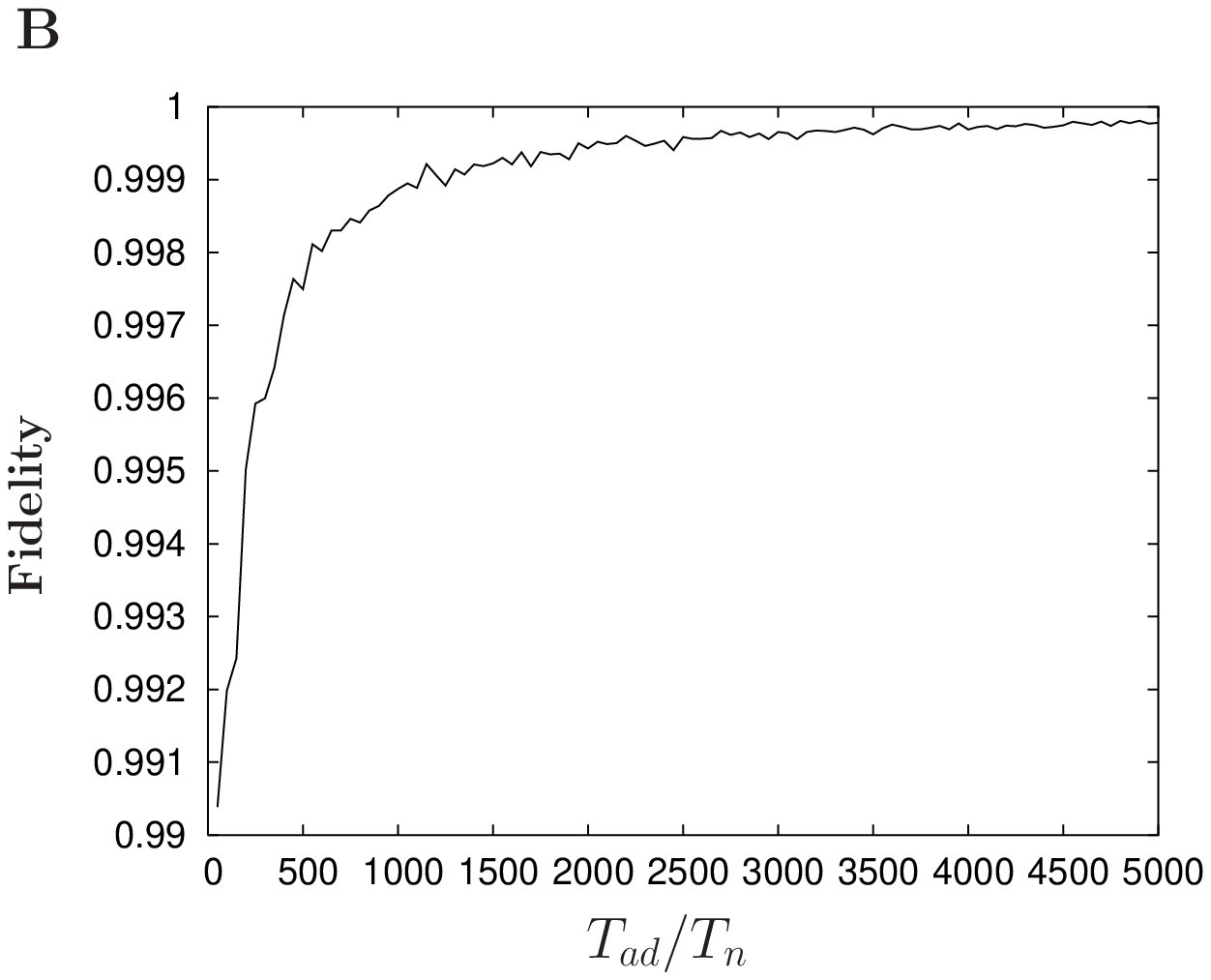}
    \caption{\label{fig:ad_int02_150_per} {\it Fidelity} for small
      perturbations $\delta \Omega =0.01 \Omega$. Parameters as in figure 
      \ref{fig:ad_int02_150}. }
  \end{center}
\end{figure}

\begin{figure}[t]
  \begin{center}
    \includegraphics[height=5cm]{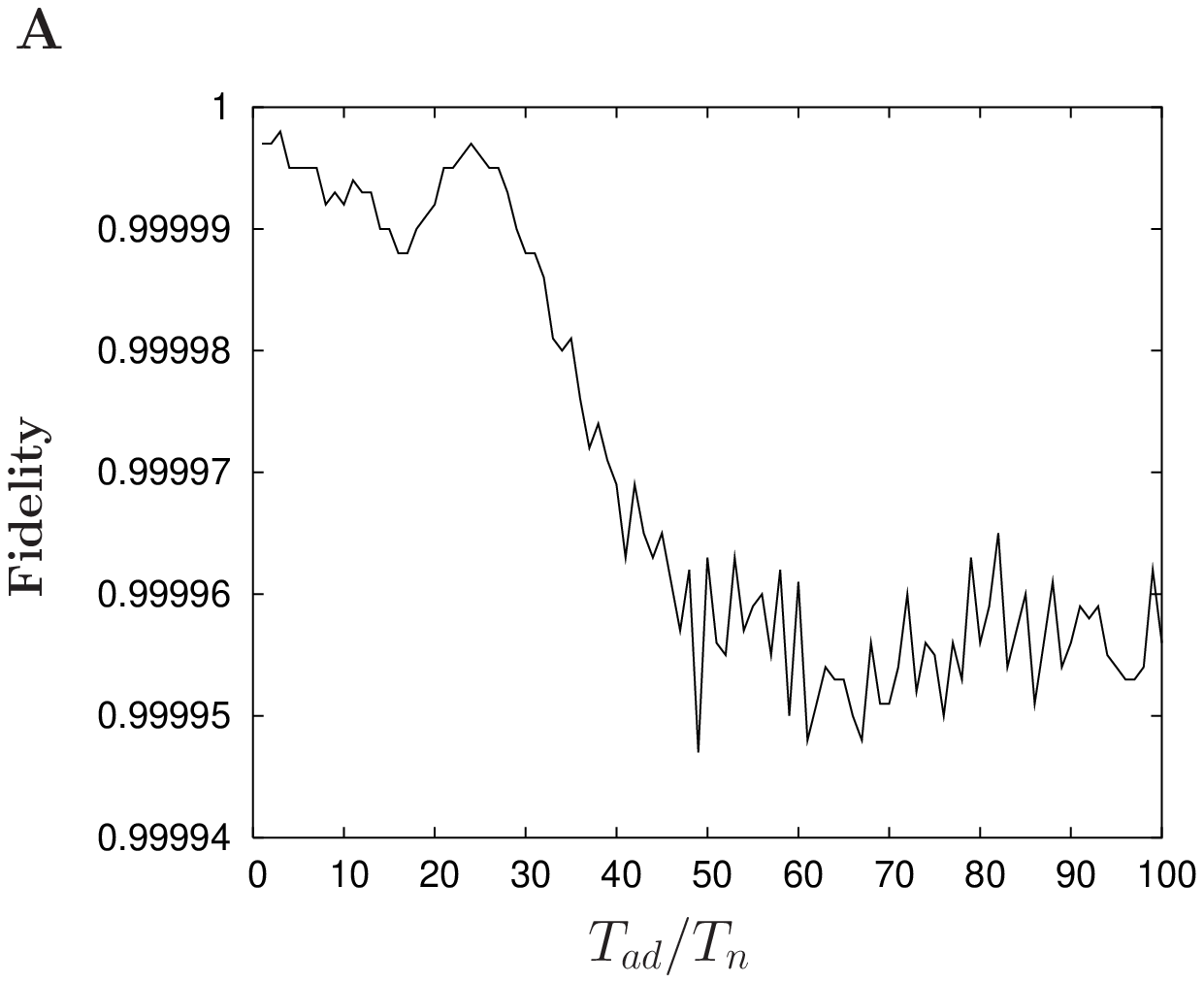}
    \includegraphics[height=5cm]{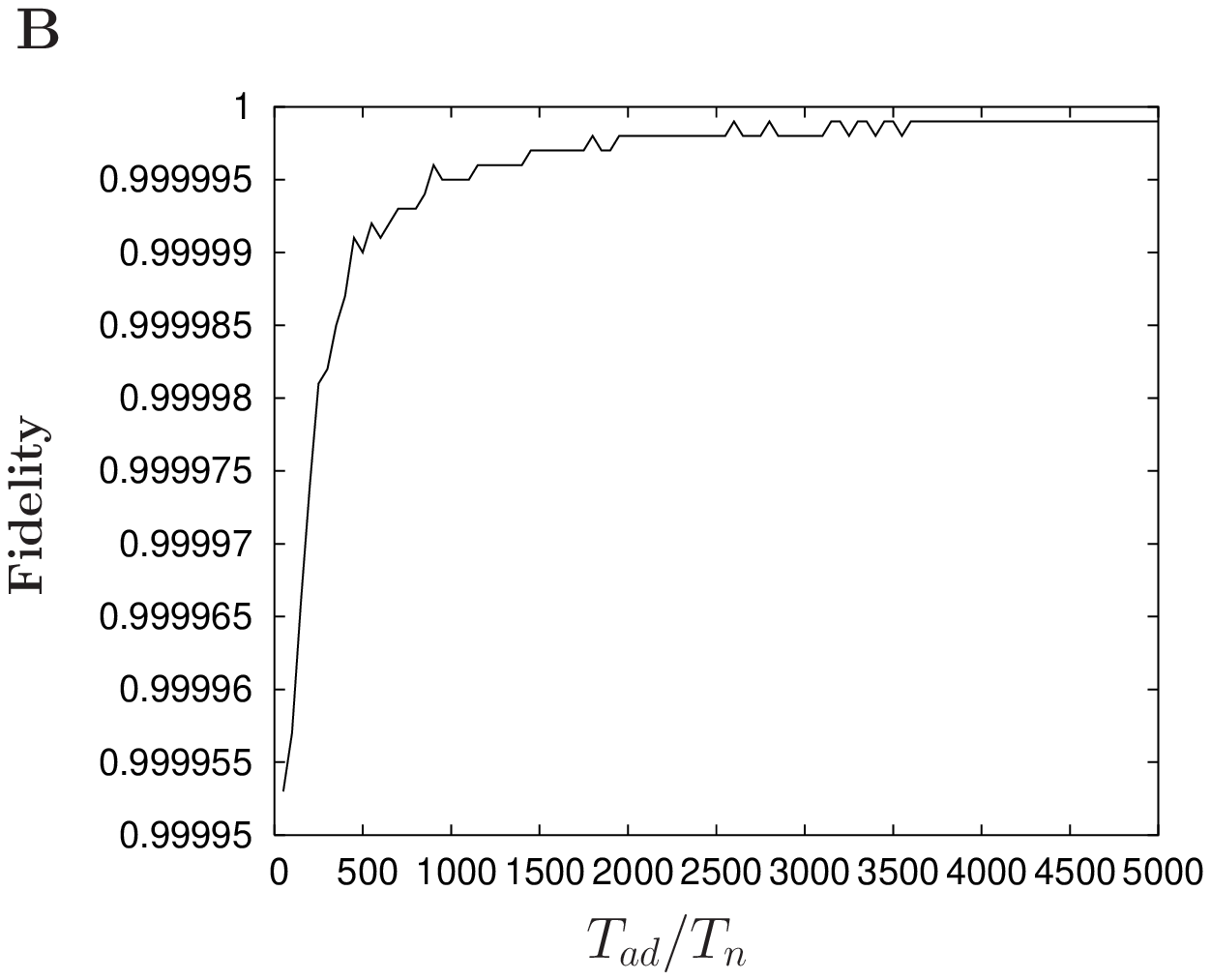}
    \caption{\label{fig:ad_ph02_150}{\it Fidelity} for the 'phase'
      noise. Parameters as in figure \ref{fig:ad_int02_150}.}
  \end{center}
\end{figure}

\begin{itemize}
  \item{} Slowly varying random  fluctuations ($T_{ad}/T_n \approx1$): 
    the loop basically maintains its shape and  it is simply shifted. 
    This situation does not affect the gate too much.
  \item{} Intermediate regime ($50 \le T_{ad}/T_n \le 100$): 
   the intense fluctuations  badly  modify modify the loop shape  and alter the gate operator.
 \item{}  Fast varying random  fluctuations
   ($T_{ad}/T_n \gg 1$):  the fluctuations effectively 
   cancel out and do not change the operator.
\end{itemize}

These geometrical features are independent from the {\it ratio}
$\delta \Omega / \Omega$ and persist even for small values of the fluctuation 
$\delta \Omega$.
If we decrease the intensity of the noise our adiabatic gate improves 
as shown in Figures (\ref{fig:ad_int02_150_per}) where, 
with the same parameters (i.e. Rabi frequency and adiabatic time), we choose 
$\delta \Omega =0.01 \Omega$.
We note that the {\it fidelity} increases but the features are the same of 
the previous simulations (i.e. three regimes).

\begin{figure}[t]
  \begin{center}
    \includegraphics[height=5cm]{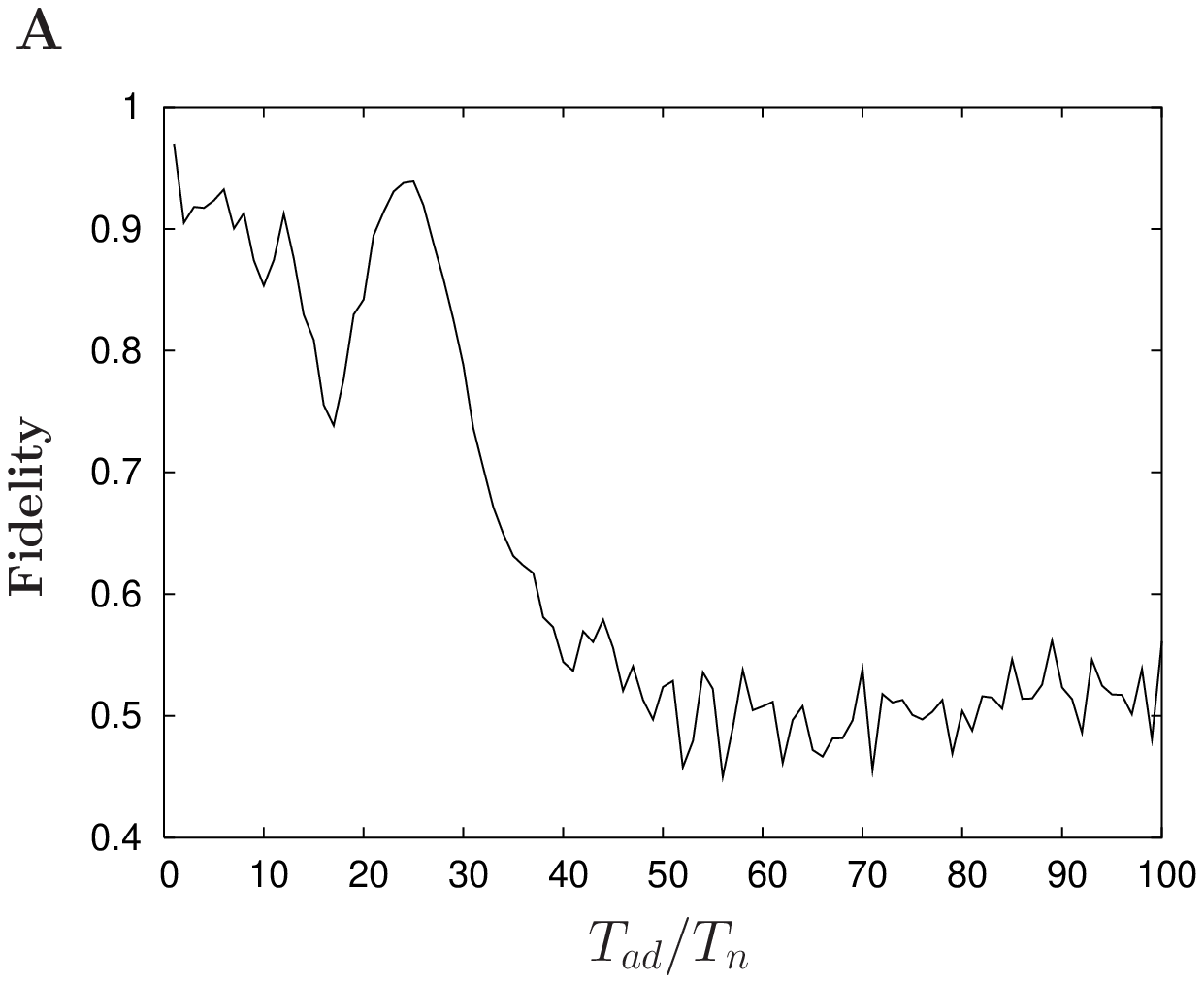}
    \includegraphics[height=5cm]{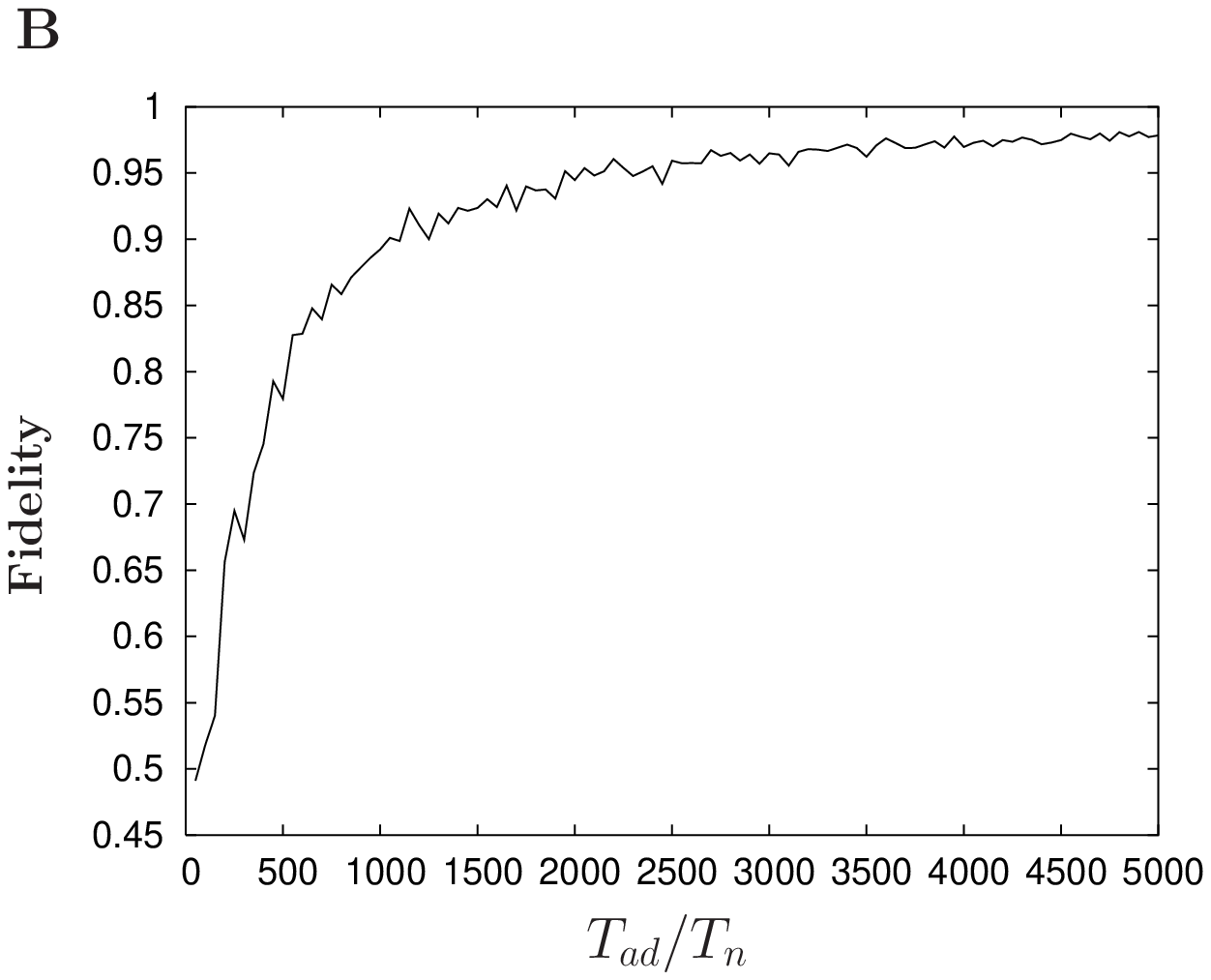}
    \caption{\label{fig:ad_both_confr}{\it Fidelity} for holonomic gates
      when the system is subject to 'intensity' and  'phase' noise.
      Parameters as in figure \ref{fig:ad_int02_150}.}
  \end{center}
\end{figure}

In Figure (\ref{fig:ad_ph02_150}) 
the noise is applied only to the phase of the control field with 
the same adiabatic parameters of the previous simulations.
Since the way of producing the operator is always the same (i.e. loop in the 
parameter space with a noise),
we expect that the effects of noise are the same 
of those for 'intensity' noise. 
This can be clearly seen in figures (\ref{fig:ad_ph02_150}) where we find 
the same features of the previous plots.
The { \it fidelity} is much better respect to 'intensity'
noise and we can say the the 'phase' noise does not affect our gate.
 
We finally discuss the case of  both intensity and phase noise 
(Figure \ref{fig:ad_both_confr}). As we expect, the main part of the error 
is given by the intensity noise .
The three regimes, discussed previously, are evident also in this case.

\begin{figure}[t]
  \begin{center}
    \includegraphics[height=5cm]{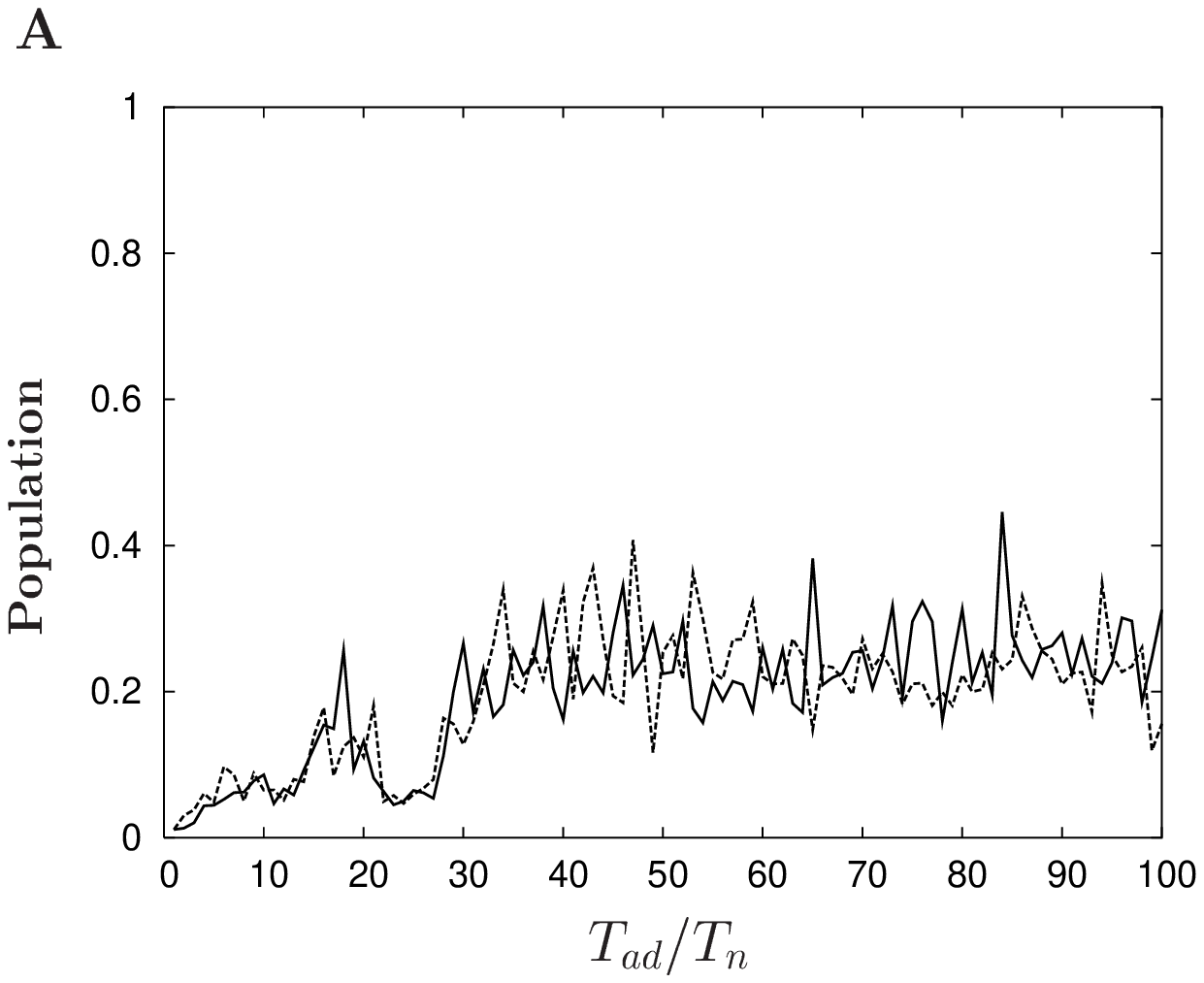}
    \includegraphics[height=5cm]{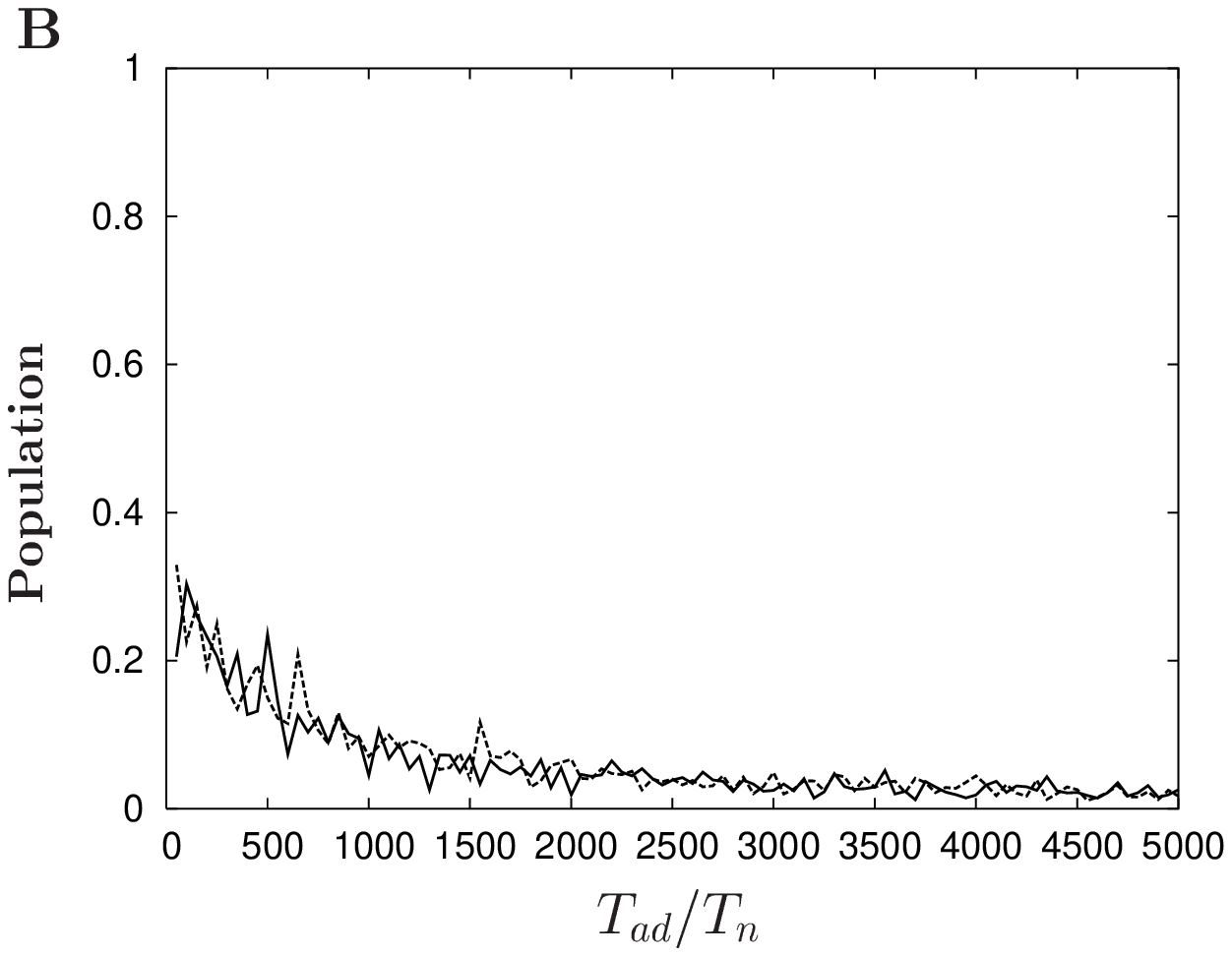}
    \caption{\label{fig:population} Population of the non-logical sates
      $|G\rangle$ and $|E^0_L\rangle$ after the holonomic gate application 
      when the system is subject to 'intensity'.
      Parameters as in figure \ref{fig:ad_int02_150}.}
  \end{center}
\end{figure}

In Figures (\ref{fig:population}) we show the population of the non-logical 
states ($|G\rangle$ and $|E^0_L\rangle$) at the end of the gate application 
as function of $T_{ad}/T_n$. In an ideal adiabatic gate (for 
$T_{ad}\rightarrow \infty$) these states are not populated.
In Figure (\ref{fig:population}) (A) these populations increase with the 
$T_{ad}/T_n$ ratio due to the strong and fast fluctuations of the Rabi 
frequencies which perturb the Hamiltonian.
For very fast varying fluctuations (Figure \ref{fig:population} (B)) the 
undesired populations decrease because of cancellation effects.
From this analysis we can say that for slowly and fast varying random 
fields (i.e. the two extreme regimes) we remain in the logical computational 
space.

We wish to compare this holonomic gate with a standard dynamical gate,
the latter one being characterized by the same unitary operator of 
the holonomic one and having different (typically shorter) 
gating time (the time need for the application of the gate).
In our system this means to consider as logical states 
$|G\rangle= |0\rangle $ and  $|E^i_L\rangle = |1\rangle$ and apply a 
$\pi-$pulse laser sequence to produce a transition 
$|G\rangle \leftrightarrow |E^i_L\rangle$.

\begin{figure}[t]
  \begin{center}
    \includegraphics[height=5cm]{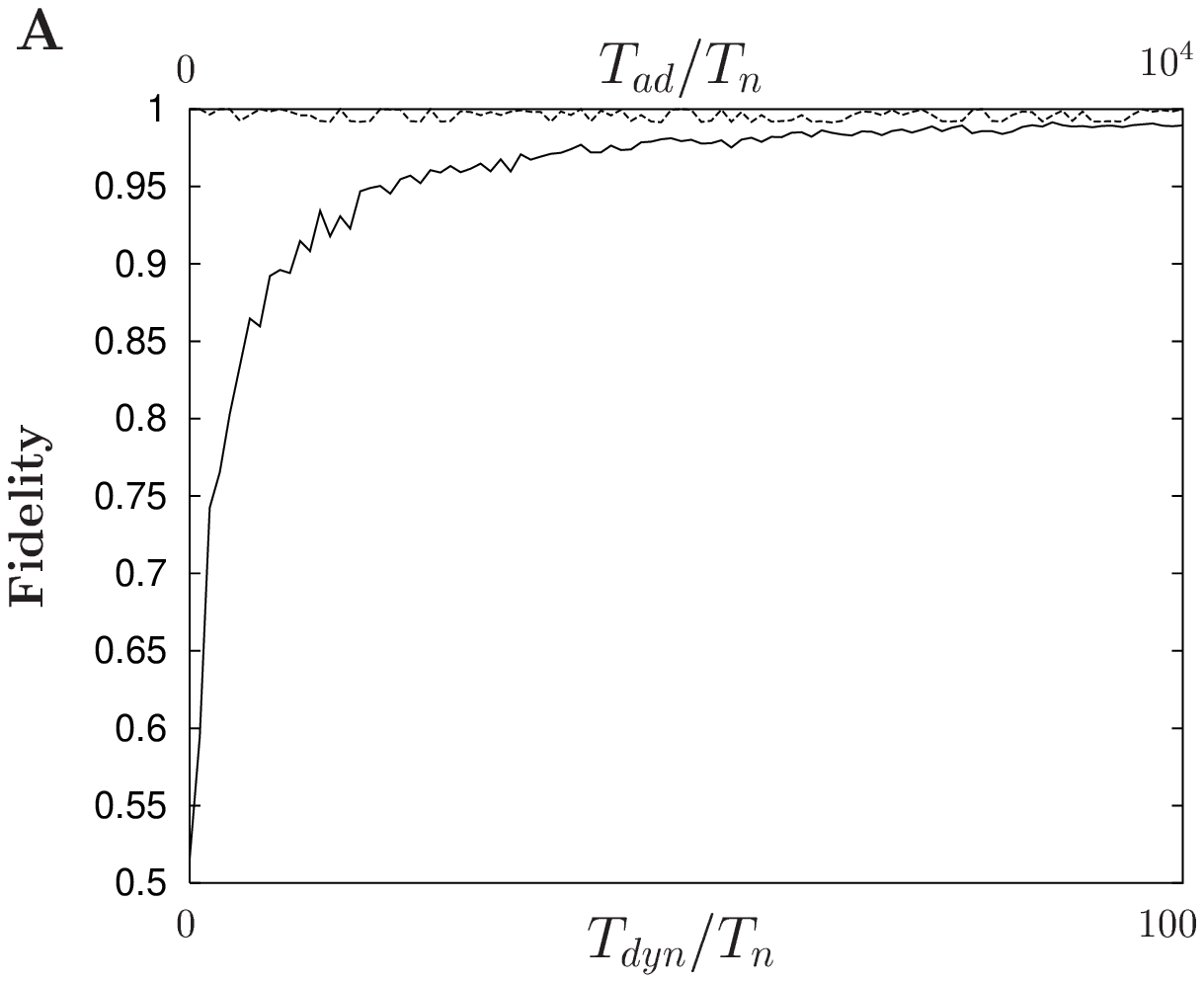}
    \includegraphics[height=5cm]{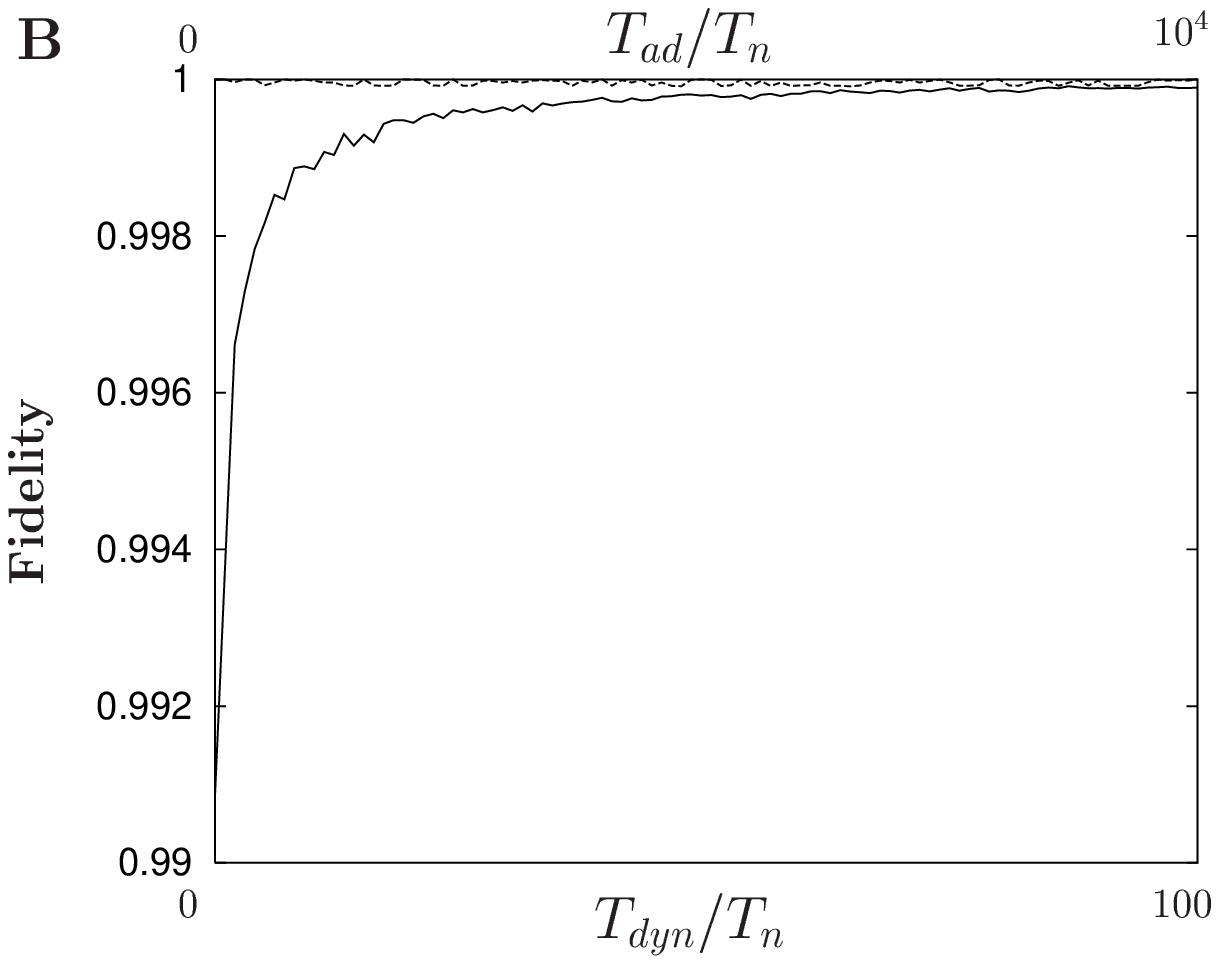}
    \caption{\label{fig:ad_dyn} Comparison between holonomic (solid line)
      and dynamical (dashed line) gates with (A) $\delta \Omega = 0.1\Omega$~ 
      and (B) with  $\delta \Omega = 0.01\Omega$ .      
      On the top $T_{ad}/T_n$ for the holonomic gate 
      and on the bottom $T_{dyn}/T_n$ for dynamical gates are reported.}
  \end{center}
\end{figure}

We made similar simulations for dynamical gate with the same parameter
($\Omega=0.02$~fs$^{-1}$) and with the same noise.

In Figures (\ref{fig:ad_dyn}) we make a comparison between holonomic 
(solid line) 
and dynamical gates (dashed line) subject to the same intensity and phase 
noise.
The comparison is not direct since the gating times  are different.
This means that the {\it ratios} $T_{ad}/T_n$ and $T_{dyn}/T_n$ are different 
for adiabatic and dynamical gates.
To compare the effect of the gate subject to the {\it same} noisy field
(i.e. with the same noise time $T_n$), we have to take into account 
that the dynamical gates, in our model, 
are about 100 times faster than the adiabatic ones
(see Refs. \cite{paper1, long_pap}).
This means that if during the adiabatic evolution the noise changes 
$n_r^{ad} = T_{ad}/T_n$ times, for the dynamical gates it  changes 
only $n_r^{dyn} = T_{ad}/(100~ T_n)$ times (if for the dynamical noise 
$n_r^{dyn} < 1$, it changes only once).
In other words, if for dynamical gate we extract  $n_r^{dyn}$ random numbers
during the evolution, for the holonomic gate we extract 
$n_r^{ad} = 100~n_r^{dyn}$ random numbers.
This means that to compare the {\it fidelities} 
we have to look in the fast varying fluctuating noise ($T_{ad}/T_n \gg 1$)
region of the previous figures.

\begin{figure}[t]
  \begin{center}
    \includegraphics[height=5cm]{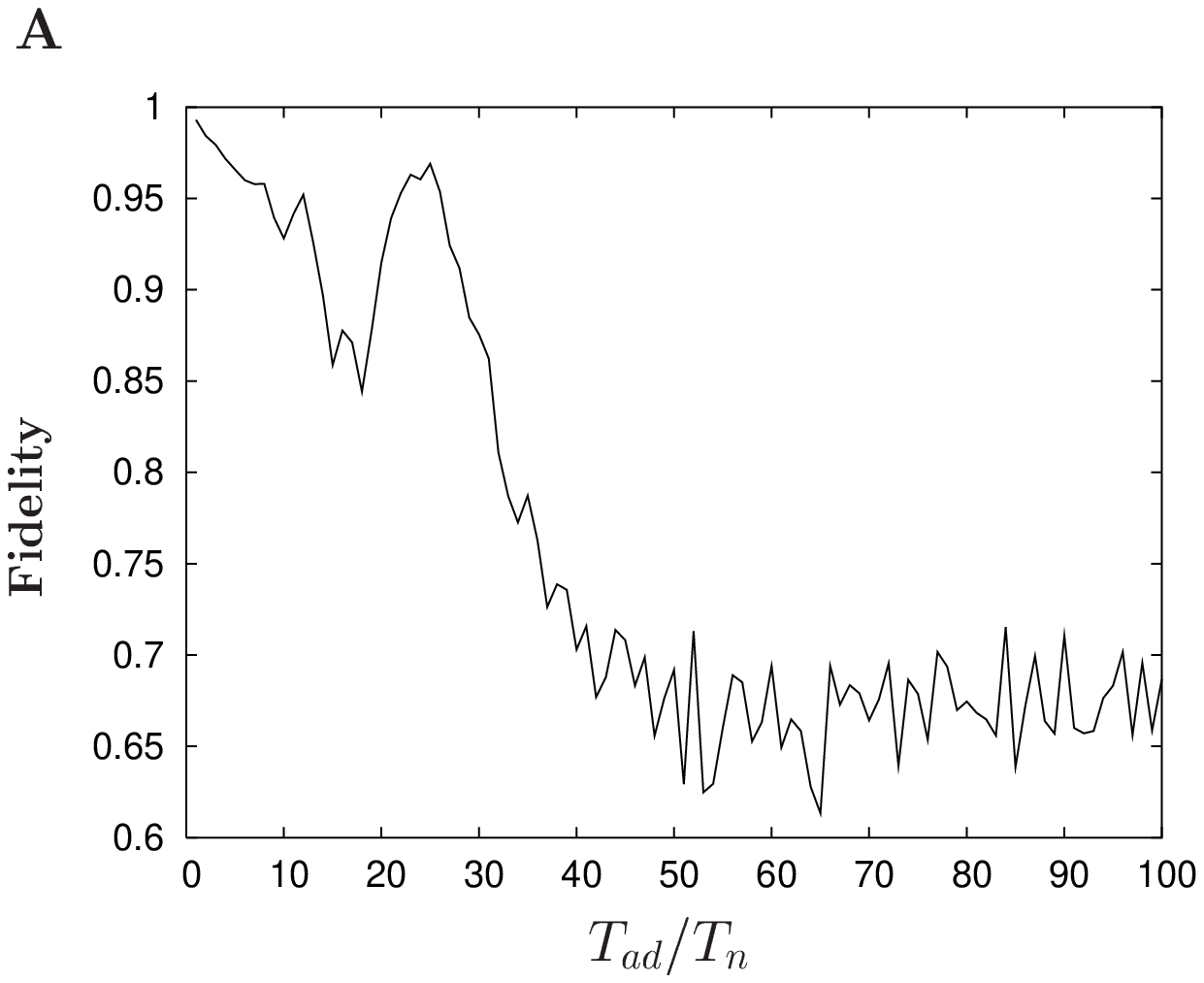}
    \includegraphics[height=5cm]{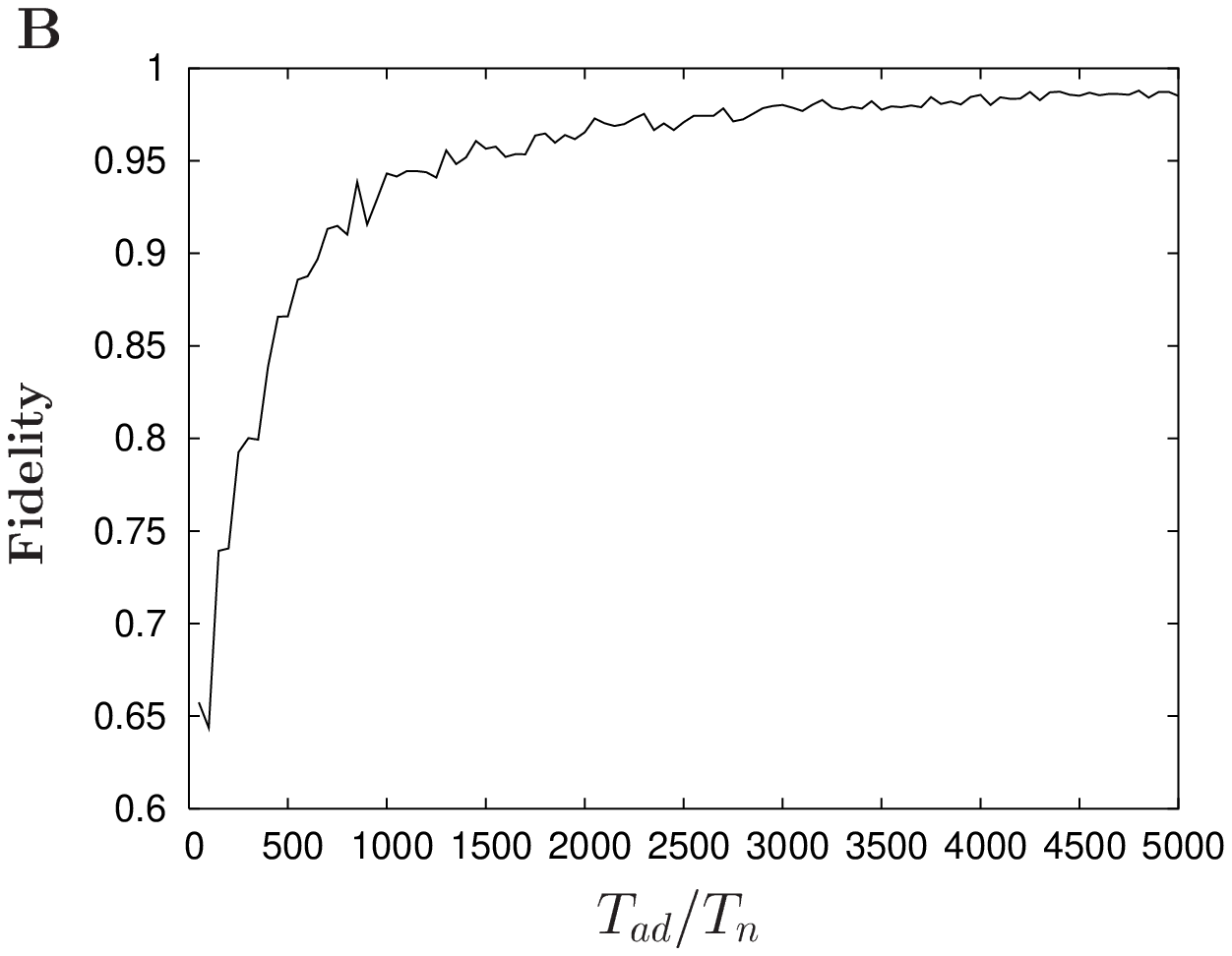}
    \caption{\label{fig:selective_phase}
      {\it Fidelity} for the single qubit {\it phase shift} gate with 
      to intensity noise.
      The parameter are the same of figure \ref{fig:ad_int02_150}}.
  \end{center}
\end{figure}

\begin{figure}[t]
  \begin{center}
    \includegraphics[height=5cm]{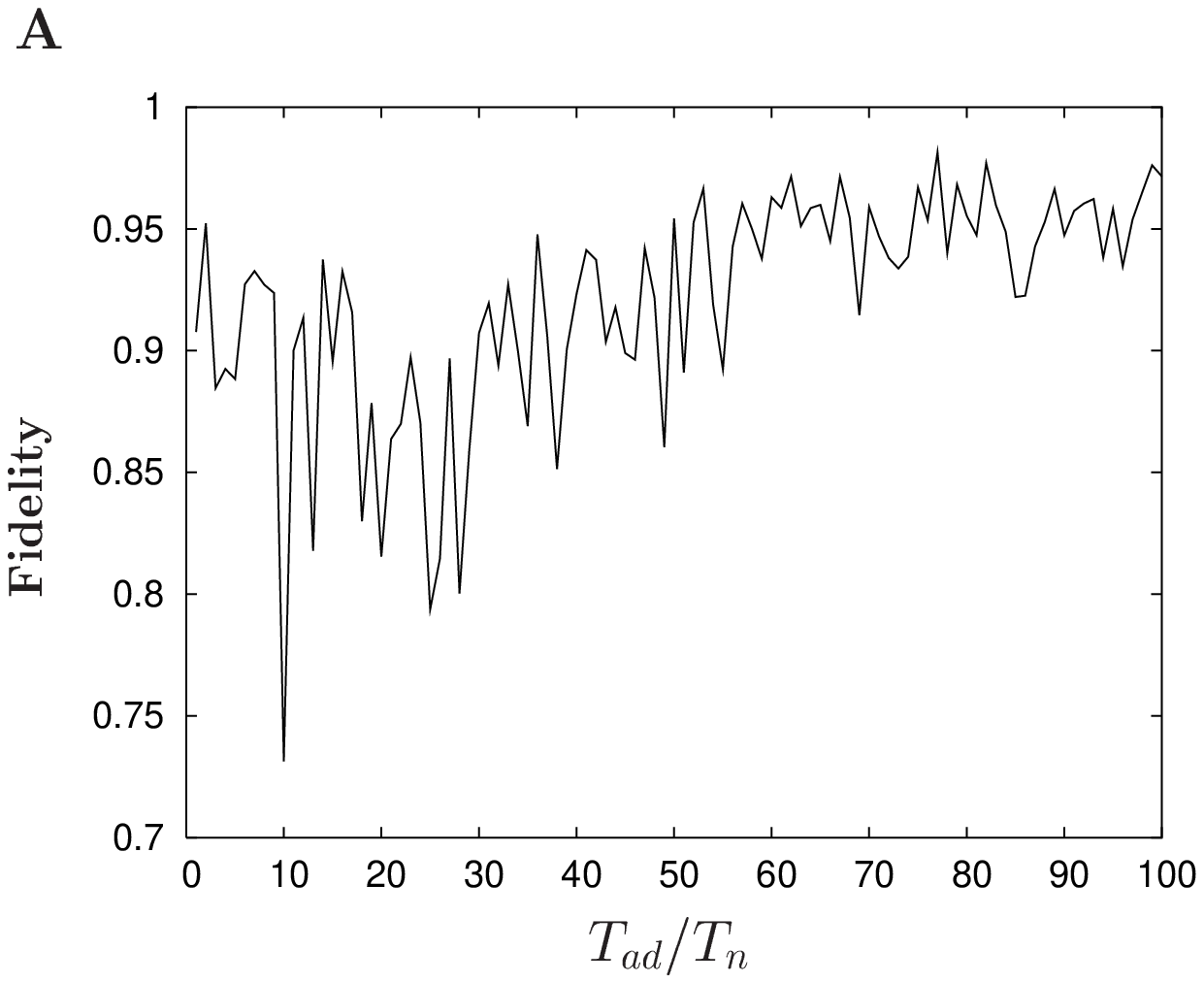}
    \includegraphics[height=5cm]{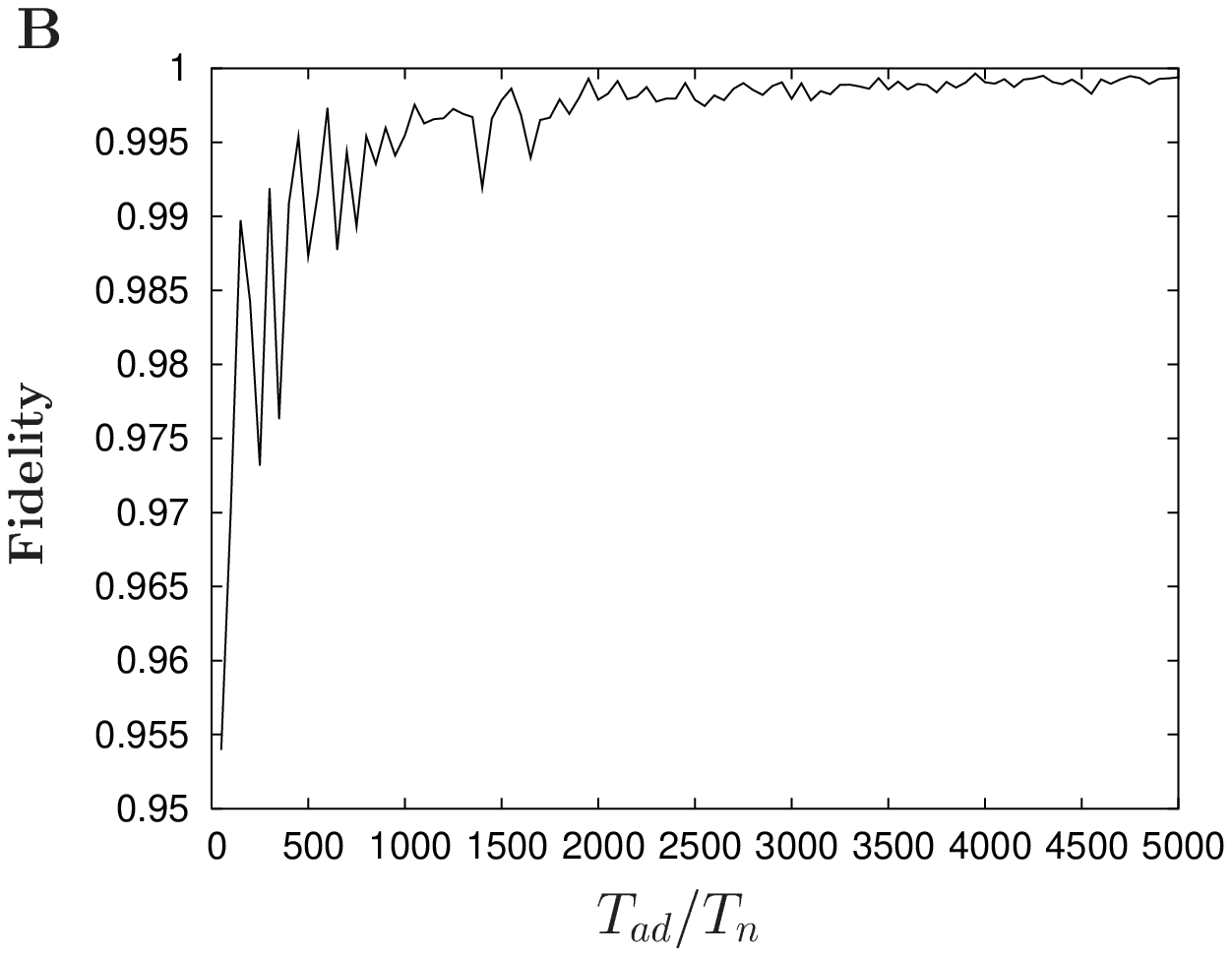}
    \caption{\label{fig:two_qubit_gate}
      {\it Fidelity} for the two qubit {\it phase shift} gate with 
      to intensity noise. Parameters are : $\delta = 5$~meV,
      $|\Omega_{i}| = \delta/15$ and $T_{ad} = 0.8$~ns.}
  \end{center}
\end{figure}

For this reason in Figures \ref{fig:ad_dyn} (A) and (B) we put two 
different scales for the adiabatic ($T_{ad}/T_n$) and dynamical ($T_{dyn}/T_n$)
gates and the plots have been rescaled taking into 
account for the different gating time.


Numerical simulations in Fig. \ref{fig:ad_dyn} show that the performance of 
holonomic gates and dynamical gates are comparable {\it in the region} 
where the first ones are reliable; that when $T_{ad}/T_n \gg 1$
(see discussion above).

Both dynamical and holonomic gates can be further improved.
Since the dynamical gates are not subject to adiabatic constraints,
we can choose different parameters 
in order to minimize the effect of the noise but this can affect the gating 
time.

For holonomic gates, given a noise with fixed correlation noise time $T_n$,
we can try to change adiabatic time in order to modify the $T_{ad}/T_n$ 
{\it ratio}.
This should allow us to fall in a 'good' regime i.e. 
fast or slowly varying fluctuations.
Decreasing adiabatic time to get in the small $T_{ad}/T_n$ region can produce
new errors due to the lack of adiabaticity during the evolution and then it
must be treat carefully.
Increasing the adiabatic time to enter in the region of great $T_{ad}/T_n$ 
leads both to a better precision in the adiabatic gates and 
to the cancellation of noisy fluctuations but gives longer gating times.

To complete the set of universal quantum gates, we need another single 
qubit gate and a two qubit gate.

In the first case, we apply an intensity noise to the {\it phase gate} 
presented in Ref. \cite{paper1} with $\Omega_{-} = 0$,
$\Omega_{+} = -\Omega \sin(\theta /2) ~e^{i \varphi}$ and
$\Omega_{0} = \Omega \cos(\theta /2)$. The holonomic operator is 
$U= e^{i \phi |E^{+}\rangle\langle E^{+}|}$ where 
$\phi = \frac{1}{2} \oint \sin\theta d\theta d\varphi$.
The results are shown in Figures (\ref{fig:selective_phase}) and 
the structures discussed above are evident.

Finally, in Figures (\ref{fig:two_qubit_gate})
we present the simulations of the {\it phase shift} two qubit gate in 
Ref. \cite{paper1} with an intensity noise. In this case we use two exciton
states ($|E^i_L E^j_L\rangle$ and $|GG\rangle$) and a two 
photon interaction Hamiltonin similar in structure to (\ref{eq:ham_ec}) 
but with Rabi frequency $\Omega_{eff} = 2 \hbar \Omega^2 / \delta$
(where $\delta$ is the laser detuning we need to avoid single photon 
transition and create two exciton states).

In Figure (\ref{fig:two_qubit_gate}) (A) 
where the second regime (where the {\it fidelity} decreases) 
is present but less evident and seems to be compressed in the slowly
varying random fluctuations zone ($10\le T_{ad}/T_n \le 30$); 
for great $T_{ad}/T_n$ ratio the 
{\it fidelity} decisively increases as in the previous figures
(Figure (\ref{fig:two_qubit_gate}) (B)).
Moreover, we note that the {\it fidelity} between the adiabatic final states 
with and without noise is high and close  to $1$ even if we have chosen the 
adiabatic parameter ($\Omega_{eff}~T$) smaller than the one used for single 
qubit gates. These can be explained as consequence of the fact that
in the imperfect adiabatic evolution unwanted states gets populated.
On the other hand  the effect of the fluctuating noise is to induce
undesired transition as well.
These two effects are superposed and with smaller adiabatic parameter 
the effect of the noise seems to be less important.

\section{Conclusions}
\label{sec:conclusions}

We numerically studied the robustness of a non-abelian holonomic quantum gate
against stochastic  errors in control parameters.
The robustness of logical gates show three regimes upon
 the variations of the noise correlation time $T_n$.
A possible interpretation of
these regimes can be given on the basis  of the  the geometric 
(i.e. solid angle swept in the 
parameter space) dependence of the holonomic operator.
For fast random varying  fluctuations we have a good robustness of the 
holonomic gate since,  as argued in other papers \cite{preskill,ellinas}, 
the fluctuations during the loop tend to cancel out.
For random varying  fluctuations in the intermediate  regime the holonomic 
gates are significantly corrupted  because the fluctuations strongly 
deform the parameter loop.
For slowly random varying  fluctuations the performance improves again. 
This fact is not surprising as it may seem, indeed the loop in the parameter 
space turns out in this case to be simply is shifted rather than deformed;  
then similar solid angles are swept.
Our analysis suggests that the main noise source is given by  fluctuations in 
the intensity of the control  parameters whereas  the phase fluctuations do 
not seem to sizeably  affect the gate studied. The effect of the noise 
decreases with the variance of the intensity of the  fluctuations and 
for $\delta \Omega/ \Omega =0.01$  we have average {\it fidelity} 
close to $1$.
A first comparison shows that holonomic and dynamical gates have comparable
performance in the $T_{ad}/T_n\gg 1$ region.

We performed similar simulations for two single qubit gates and for 
a two qubit gate in order to complete the set of universal quantum gates.
For the single qubit gates we obtain similar results.
For the two qubit gate the three regimes descibed above are present but less 
evident.

We believe that our analysis and conclusions should be extended rather  
easily to different sort of system proposed for HQC:
for example, it  definitely extends to the model  proposed in ref. \cite{duan}
since the involved holonomic structure is isomorphic  to the one here analyzed.
The general features of our results  should hold  also more general situations,
 since  they apparently  do not rely on the detailed features of Hamiltonian 
(\ref{eq:ham_ec}), rather on the general structure of holonomic 
computations. A related, though logically distinct, issue is the robustness
of HQC against environmental decoherence \cite{florian}.
This is clearly an important subject to be addressed in future investigations.

P.Z. gratefully acknowledges financial support by Cambridge-MIT Institute Limited and by the 
European Union project  TOPQIP (Contract IST-2001-39215)


\begin{thebibliography}{99}
\bibitem{crypto} 
 C.H. Bennett and G. Brassard {\it Proceedings of IEEE International 
   Conference on Computers, Systems and Signal Processing} 175-179,
   IEEE, New York, 1984.
   C. H. Bennett {\it Phys. Rev. Lett.} {\bf 68}(21), 3121, 1992. 

\bibitem{teleportation} 
  C.H. Bennett G. Brassard, C. Crepeau, R. Jozsa, A. Peres, and 
   W. K. Wootters Phys. Rev. Lett. {\bf 70}, 1895, 1993.

\bibitem{algos} P.W. Shor {\it Proceeding of 35th Annual symposium on 
  foundation of Computer Science}. 
  (IEEE Computer Society Press, Los Alamitos, CA, 1994).
  L.K. Grover {\it Proc. $28^{th}$ Annual ACM Symposium on the Theory of 
    Computation}, 212, ACM Press, New York, 1996.
\bibitem{error_avoiding}
  P. Zanardi and M. Rasetti,  
  Phys. Rev. Lett. {\bf 79}, 3306 (1997).
\bibitem{error_correcting}
  P.W. Shor Phys. Rev.A {\bf 52}, 2493 (1995);
  A.M. Steane Phys. Rev. Lett. {\bf 77}, 793 (1996);
  E. Knill,  R. Laflamme, 
  Phys. Rev.A {\bf 55}, 900 (1997) and references therein.
\bibitem{bang-bang}
L. Viola and S. Lloyd, Phys. Rev. A. {\bf 58}, 2733 (1998);
L. Viola, E. Knill, and S. Lloyd Phys.Rev.Lett. {\bf 82},  2417 (1999);
D. Vitali, P. Tombesi, Phys. Rev. A {\bf 65}, 012305 (2002);
P. Zanardi, Phys. Lett. A {\bf 258} 77 (1999)

\bibitem{topological}
  A. Kitaev , Preprint quant-ph/9707021.
  M.H. Freedman, A.  Kitaev, W.  Zhenghan,
  Commun.Math.Phys. {\bf 227} 587 (2002).

\bibitem{tns} P. Zanardi, S. Lloyd, Phys. Rev. Lett. {\bf 90}, 067902 (2003)

\bibitem{HQC}
  P. Zanardi and M. Rasetti, Phys. Lett.A {\bf 264}, 94 (1999).
  J. Pachos, P. Zanardi and M. Rasetti,
  Phys. Rev.A {\bf 61}, 010305(R) (2000).

\bibitem{abelian}
  J.A. Jones  {\it et al.} , Nature  {\bf 403}, 869 (2000).
  G. Falci  {\it et al.} , Nature  {\bf 407}, 355 (2000).

\bibitem{HQC_proposal}
   R.G. Unanyan, B.W.  Shore and  K. Bergmann,
   Phys. Rev. A {\bf 59}, 2910 (1999).
   L. Faoro, J. Siewert and R. Fazio, Phys. Rev. Lett. {\bf 90}, 028301 (2003).
   I. Fuentes-Guridi {\it et al.} Phys. Rev. A {\bf 66}, 022102 (2002).
   A. Recati {\it et al.} Phys. Rev. A {\bf 66}, 032309 (2002).
                                                                               
\bibitem{duan}
   L.-M. Duan,J. I.  Cirac and P. Zoller, Science {\bf 292}, 1695 (2001).
                                                                             
\bibitem{paper1}
P. Solinas {\it et al.}
Phys. Rev. B {\bf 67}, 121307 (2003)
                                                                               
\bibitem{long_pap}
P. Solinas {\it et al.}
Phys. Rev. A {\bf  67}, 062315 (2003)
                                                                              
\bibitem{non_adiabatic}
  WangXiang-Bin, M. Keiji  Phys. Rev. Lett. {\bf 87},  097901 (2001); 
  WangXiang-Bin, M. Keiji   Phys. Rev. Lett. {\bf 88}, 179901(E) (2002).
  X.-Q.  Li {\it et al.} Phys. Rev. A {\bf 66}, 042320 (2002).
  S. L. Zhu, Z.D. Wang, Phys. Rev. Lett. {\bf 89}, 097902 (2002).
  P.Solinas {\it et al.} Phys. Rev. A {\bf 67}, 052309 (2003)

\bibitem{preskill} J. Preskill in {\it Introduction to Quantum Computation
        and Information}, edited by H.-K. Lo, S. Poposcu, and T. Spiller
        (World Scientific, Singapore, 1999).
                                                                               
\bibitem{ellinas} D. Ellinas and J. Pachos,  Phys. Rev. A {\bf 64},
022310 (2001).

\bibitem{HQC_noise}
 A. Blais and A.-M. S. Tremblay
 Phys. Rev. A 67, 012308 (2003);
 A. Nazir, T. P. Spiller, and W. J. Munro, Phys. Rev. A 65, 042303 (2002);
 G. De Chiara, G.M. Palma, Phys. Rev. Lett. {\bf 91}, 090404 (2003);
 A. Carollo {\it et al}, Phys. Rev. Lett. 90, 160402 (2003);
 A. Carollo {\it et al}  quant-ph/0306178;
 V.I. Kuvshinov, A.V. Kuzmin, Phys. Lett. A, {\bf 316}, 391 (2003);
 F. Gaitan, quant-ph/0312008  

\bibitem{W-Z} F. Wilczek ,A. Zee,
{\it Phys. Rev. Lett.} {\bf {52}}, 2111 (1984).

\bibitem{sample}
Our sampling set of the Bloch sphere is given by
$\{ \pm e_i\}_{i=x,y,z}, (\pm e_x\pm e_y)/\sqrt{2}, 
(\pm e_x\pm e_z)/\sqrt{2}, (\pm e_y\pm e_z)/\sqrt{2}$.
Here $e_i$ denotes the normalized vector of the $i$-th direction.

\bibitem{florian}
I. Fuentes-Guridi, F. Girelli, E. R. Livine,
 quant-ph/0311164
\end{thebibliography}
\end{document}